\documentstyle[epsfig]{aipproc}

\newcommand{\postscript}[2]
{\setlength{\epsfxsize}{#2\hsize}
\centerline{\epsfbox{#1}}}

\begin{document}
\title{Precision Measurements\\  at The Higgs Resonance: \\
A Probe  of Radiative Fermion Masses\thanks{Talk presented
by N. Polonsky at the {\it Workshop on
Physics at the First Muon Collider and the Front End of a Muon 
Collider}, Fermilab, November 6-9, 1997.
Work supported by Schweizerischer Nationalfonds
and by the  US National Science Foundation grant No.~PHY-94-23002.} }
\author{F. M. Borzumati$^a$, G. R. Farrar$^b$,
N. Polonsky$^{b}$, and S. Thomas$^c$}
\address{$^a$ Institut f\"ur Theoretische Physik,
Universit\"at Z\"urich,
CH - 8057 Z\"urich, Switzerland\\
$^{b}$ Department of Physics and Astronomy, Rutgers University, 
Piscataway, NJ 08854, USA\\
$^{c}$ Physics Department, Stanford University, Stanford, CA 94305, USA}

\hfill \begin{flushright}
ZU-TH 35/97\\
RU-97-96\\
SU-ITP 97-52\\
\vspace*{1cm}
\end{flushright}
\maketitle

\begin{abstract}
The possibility of radiative generation of fermion masses 
from  soft supersymmetry breaking chiral flavor violation 
is explored.
Consistent models are identified and
classified.  Phenomenological implications for electric dipole moments and
magnetic moments, as well as  collider probes --  in particular those
relevant at the Higgs resonance --  are discussed. It is shown that 
partial widths  $\Gamma_{h^{0} \rightarrow ff}$ are enhanced
compared with the minimal supersymmetric standard model.
\end{abstract}

\newpage

\section*{Introduction}

The main motivations for proposing and building 
the next generation(s) of hadron and lepton colliders
are the discovery of the Higgs boson and
the search for signals of physics beyond the Standard Model (SM).
Most of the related theoretical work has focused
so far on discovery strategies and on the extraction
of various parameters within a given theoretical framework.
These issues have been and are still extensively addressed
in the framework of supersymmetry, in general, and
the minimal supersymmetric standard model (MSSM), in particular \cite{susy}.
Recently, it has been shown that electron colliders
could efficiently probe various quantum corrections 
in supersymmetric models.
One-loop corrections to gaugino couplings, 
that grow logarithmically with the scale of
heavy (and kinematically unaccessible)  sparticles,
modify observables involving light sparticles and may be measured \cite{sob}.
Hence, if supersymmetry is present at low-energies
one may gain a more complete understanding of its realization 
in a similar manner to the knowledge gained from precision probes of
quantum corrections in the electroweak theory.

Complementary measurements 
(in lepton flavor and energy range) 
of the corrections discussed above are also
possible in a muon collider \cite{vb}.
Here, however, we would like to consider
a unique feature of the muon collider -- its
beam energy resolution which enables it to explore narrow resonances. 
In particular, an $s$-channel Higgs boson \cite{schannel}
could be produced and its properties studied.
The resonant Higgs production offers an opportunity to explore quantum
effects in the Higgs sector of supersymmetric models.
Most studies of the Higgs resonance have focused so far 
$(i)$ on distinguishing the SM Higgs boson from the
light supersymmetric Higgs boson, and 
$(ii)$ on extracting the properties of the heavier Higgs bosons
in supersymmetric and two Higgs doublet models \cite{higgs}.
Here, we propose that precision
measurements of a partial width $\Gamma_{H \rightarrow ff}$,
along with other low-energy observables discussed below, 
can reveal
whether the respective fermion mass is  radiative \cite{rad,paper}
with a corresponding ``soft'' (rather than  tree-level) Yukawa coupling.
If the Yukawa couplings are induced by supersymmetry breaking quantum effects,
they would enable one to indirectly probe the supersymmetry breaking sector.
We note in passing that even if there are tree-level Yukawa couplings,
the quantum corrections in supersymmetric models can be 
possibly measurable.

We will demonstrate how radiative scenarios
may arise in supersymmetric frameworks and explore
their implications for fermion masses and couplings.
We will show that a soft Yukawa coupling arising from supersymmetry
breaking chiral flavor violation
is always enhanced in comparison
with the case of a tree-level coupling.
In the case of the muon, however, the enhancement is constrained 
by the upper bound on the muon anomalous magnetic moment.
Before concluding, we will suggest possible avenues for more detailed
collider studies. Here, we will concentrate on the phenomenology
of the light supersymmetric Higgs boson, $h^{0}$, which
may be produced in a low-energy machine ($\sqrt{s} \leq 200 - 500$ GeV). 
Details, as well as applications involving the heavy
Higgs bosons and the Higgsinos, can be found in Ref.~\cite{paper}.

\section{Model Building}

In the absence of tree-level Yukawa couplings, chiral flavor symmetries can
still be broken by trilinear terms in the scalar potential,
\begin{equation}
V = \sum
m_{i}^{2}\phi_{i}^{2} + [B_{ij}\phi_{i}\phi_{j} + 
A_{ijk}\phi_{i}\phi_{j}\phi_{k}
+ A^{\prime}_{ijk}\phi_{i}^{*}\phi_{j}\phi_{k} + h.c.]
+\lambda_{ij}\phi_{i}^{2}\phi_{j}^{2}.
\label{V}
\end{equation}
In this case, the chiral flavor symmetries 
in the fermion sector are  broken at the quantum level.
Gauge loops $\propto A$ or $\propto A^{\prime}$ 
which dress the fermion propagator 
generate the fermion mass and its effective coupling
to the Higgs bosons. We will return to the generation of masses and
couplings below. First, however, we would like to elaborate on
the possible realizations of such a framework.

The flavor symmetries of the high-energy theory
can forbid certain fundamental Yukawa
couplings but allow for either $(i)$ $ZH\Phi_{L}\Phi_{R}/M$
or $(ii)$ $ZZ^{\dagger}H^{\dagger}\Phi_{L}\Phi_{R}/M^{3}$
(superfield) operators in the superpotential or the Kahler potential, 
respectively. The chiral superfield $Z = z + \theta^{2}F_{Z}$ 
parameterizes here the supersymmetry breaking sector,
and $\langle F_{Z} \rangle = M_{SUSY}^{2}$ signals supersymmetry
breaking at a scale $M_{SUSY}$.
If the scalar component $\langle z \rangle$ vanishes and 
the  auxiliary component $\langle F_{Z} \rangle$ does not,
then no Yukawa couplings arise but only soft supersymmetry breaking 
trilinear terms $\propto \langle F_{Z} \rangle^{n} $ in the scalar potential.
The operators $(i)$ lead to $A$-type terms, $AH\phi_{L}\phi_{R}$,
while the operators $(ii)$ lead to $A^{\prime}$-type terms, 
$A^{\prime}H^{*}\phi_{L}\phi_{R}$.
The trilinear terms are not proportional to any Yukawa couplings.
The  symmetries of the models typically allow
for only one type of operators for a given flavor,
as we will assume.
Note that a sufficiently large $A^{\prime} \sim M_{SUSY}^{4}/M^{3}$ 
requires that the supersymmetry breaking scale, $M_{SUSY}$,
and the scale that governs the dynamics in the Kahler potential,
$M$, are both relatively low-energy scales. Such a situation could arise,
for example, if there is strong dynamics at  the scale $M$.

Our results below imply
$A/m \sim m_{q}(M_{\mbox{\tiny weak}})/(1.5 - 3 {\mbox{ GeV}}),
\,\, m_{l}(M_{\mbox{\tiny weak}})/(50 - 100 {\mbox{ MeV}})$ for correct quark and lepton mass
generation, respectively, assuming  a typical sfermion mass scale $m$
(and similarly for $A^{\prime}/m$).
Hence, the maximal magnitude of the trilinear parameters that can be realized
consistently determines which fermion masses can be generated radiatively.
Their magnitude is constrained most significantly by the
requirement of a stable color and charge conserving minimum.
One has the sufficient (but not necessary) constraint (for any flavor indices)
$\left| {A_{ijk}}/{m} \right| \lesssim \sqrt{3\lambda}$,
where $\lambda = \lambda_{ij}+ \lambda_{ik}+ \lambda_{jk}$,
and similarly for $A^{\prime}/m$.
Models with $A$-type trilinear operators and with
minimal (MSSM) matter content 
have $\lambda = 0$ at tree level and $\lambda < 0$ at one loop.
Such models are  inconsistent with the stability constraint.
On the other hand, there are other viable possibilities:
$(a)$ For minimal matter content but with $A^{\prime}$-type
operators one has $\lambda = g^{\prime 2}/2 \sim 0.06 $ 
(from the $D$-term potential),
where $g^{\prime}$ is the hypercharge coupling.
For the $b$-quark one has in addition (from the $F$-term potential)
$\lambda = g^{\prime 2}/2 + y_{t}^{2} \sim 1$, but only 
for a large $t$-quark Yukawa coupling, $y_{t}$, at tree-level.
$(b)$ Mirror matter flavor breaking:
An exotic multi-TeV  sector with vector-like matter,
which is  allowed by the symmetries to mix with 
the SM matter with a typical Yukawa coupling $y$, gives
$\lambda \sim y^{2}[(m_{SB}^{2} + m_{S}^{2})/m_{SB}^{2}]
\sim {\mbox{a few}}$.
The exotic matter is assumed to have supersymmetry conserving,
$m_{S}$, and breaking, $m_{SB}$, masses of the same order of
magnitude. Integrating out the exotic matter generates the quartic 
scalar terms.
One concludes that $e,\,d,\,s,\,b,\,u,\,c$ masses
can all be realized in either case $(a)$ or $(b)$. The muon mass
(and perhaps the $\tau$ mass, where
the sfermion mixing constraint $A/m \lesssim m/\langle H \rangle$
is also relevant)
can be realized only in case $(b)$.
Realization of radiative fermion masses from trilinear terms requires
either an unconventional scenario of supersymmetry breaking,
non-minimal matter content, or both,
offering indirect probes of such scenarios.

\section{The Phenomenology}

\begin{figure}[t]
\postscript{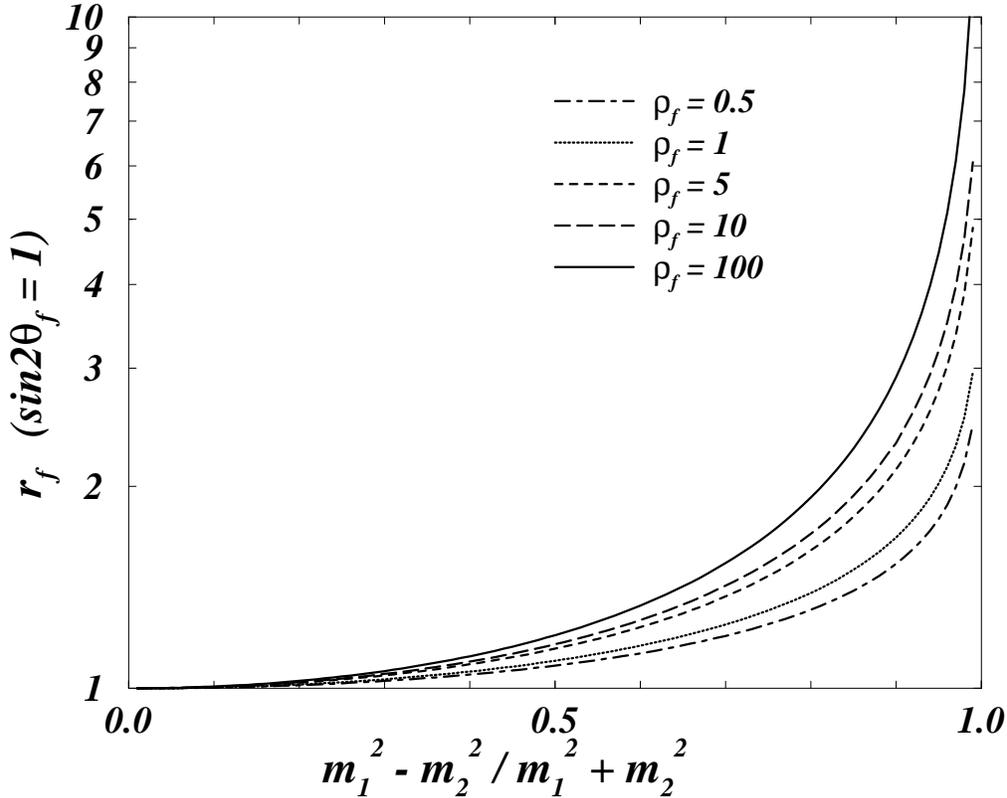}{0.85}
\caption{The ratio $r_{f}$ of the radiative soft Yukawa coupling
to a tree level Yukawa coupling for a given fermion flavor
as a function of the mass splitting 
$(m_{\tilde{f}_{1}}^{2} - m_{\tilde{f}_{2}}^{2})/
(m_{\tilde{f}_{1}}^{2} + m_{\tilde{f}_{2}}^{2})$.
The different curves correspond to different values
of $\rho_{f} = (m_{\tilde{f}_{1}}^{2} +
m_{\tilde{f}_{2}}^{2})/2m_{\lambda}^{2}$.}
\label{fig:fig1}
\end{figure}

The one-loop sfermion-gaugino exchange
which dresses the fermion propagator generates a finite contribution
to the fermion mass. It is given by
\begin{equation}
 m_f = -m_{LR}^{2}\left\{
 \frac{\alpha_{s}}{2 \pi} C_f  m_{\tilde{g}}  
  I(m^{2}_{\tilde{f}_{1}}, m^{2}_{\tilde{f}_{2}},
m^2_{\tilde{g}})  
 + \frac{\alpha^{\prime}}{2 \pi}  
m_{\tilde{B}} 
  I(m^{2}_{\tilde{f}_{1}},
m^{2}_{\tilde{f}_{2}},m^2_{\tilde{B}})
\right\},
\label{mass}
\end{equation}
where $C_{f} = 4/3, 0$ for quarks and leptons, respectively,
and $m_{LR}^{2} = A \langle H \rangle$ or $A^{\prime} \langle H^{*} \rangle$. 
The first and second terms correspond to the QCD (gluino) and
hypercharge (bino) contributions, respectively.
(Corrections due to possible neutralino mixing are omitted here).
The function $I(m^{2}_{\tilde{f}_{1}}, m^{2}_{\tilde{f}_{2}},m^2_{\lambda})$  
can be typically approximated 
$I(m^{2}_{\tilde{f}_{1}}, m^{2}_{\tilde{f}_{2}},m^2_{\lambda}) 
\times
\max(m_{\tilde{f}_{1}}^{2},m_{\tilde{f}_{2}}^{2},m^2_{\lambda})
\simeq {\cal{O}}(1)$, where $\lambda$ ($\tilde{f}$) 
denotes a gaugino (sfermion).
This leads to the numerical results given in the previous section.
Note that the radiatively generated fermion mass
does not vanish for large sparticle masses, provided that 
$A$ (or $A^{\prime}$), $m_{\tilde{f}_{i}}$, and $m_{\lambda}$,
are all of the same order of magnitude.
An effective Yukawa coupling $\bar{y}_{f}Hff$, which is momentum
dependent,  is generated by
the corresponding loop diagrams.\footnote{It differs numerically from
the Higgsino-sfermion-fermion coupling \cite{paper}.}
When applied to the decay $H \rightarrow ff$,  
it depends only on internal and external masses.
Furthermore, it is simplified for a light Higgs boson\footnote{In the case
of a massive Higgs boson the vertex is described by a $C$ function.
One also finds enhancements in this case \cite{paper}.}  
($m_{h^{0}}/m_{\tilde{f}}, \,\,
m_{h^{0}}/m_{\lambda} \rightarrow 0$) \cite{paper},
\begin{equation}
\bar{y}_{f} 
= \frac{m_{f}}{\langle H \rangle}\left\{
\sin^{2}2\theta_{\tilde{f}}\left[\frac{1}{2}
\frac{\sum_{i}I(
m_{\tilde{f}_{i}}^{2},
m_{\tilde{f}_{i}}^{2},
m_{\lambda}^{2})}{I(
m^{2}_{\tilde{f}_{1}},
m^{2}_{\tilde{f}_{2}},m^2_\lambda)}
- 1\right] + 1\right\}.
\label{vertex2}
\end{equation}  
The ratio $r_{f}\equiv \bar{y}_{f}/(m_{f}/\langle H
\rangle)$ is illustrated in Fig.~\ref{fig:fig1} 
for sfermion mixing angle $\sin 2 \theta_{\tilde{f}} = 1$,
which maximizes the effect. 
One observes that the radiative Yukawa  coupling 
could be enhanced by a significant percentage
in comparison to the case of a tree-level fermion mass.
The enhancement
increases with the mass splitting between the sfermion eigenstates.
Most importantly, we would like to stress that the soft coupling
is always enhanced.
Note that  the projecting factors between the
physical and interaction Higgs eigenstates were omitted above. 
In the case of $A^{\prime}$-type operators these factors are
different than in the usual case of tree-level couplings.
For the light Higgs boson this is irrelevant in the limit in which 
the heavy Higgs bosons decouple, and which applies to most of 
the parameter space. 

The radiative fermion mass also has strong 
implications for low-energy phenomena. 
The mass and one-loop electric dipole moment arise from 
closely related diagrams, 
implying that the phases are aligned, 
${\mbox{Arg}}(m_{e}) = {\mbox{Arg}}(d_{e}^{SUSY})$. 
The one-loop electric dipole moment therefore automatically vanishes.
Similarly, if the mass matrix of a whole sector 
({\em i.e.}. lepton, down, or up) is generated radiatively
and the soft supersymmetry breaking masses $m_{i}^{2}$ are
flavor independent, then the trilinear parameters and the
corresponding fermion masses are diagonalized  
simultaneously, suppressing potential contributions to flavor
changing neutral currents. Such a scenario is possible at least in
the case of a radiatively generated down-quark sector \cite{paper}.

The magnetic moment operator 
is also given by a loop diagram which is similar to the mass diagram.
Hence, it is not suppressed
by a loop factor compared to the radiatively generated mass
as is the case for a tree-level mass, leading to
$g_{f} -  2 \sim m_{f}^{2}/m_{\lambda}^{2}$ \cite{paper}.
If $m_{\mu}$ is radiative one predicts a muon  anomalous magnetic moment 
of the order of current limits
(${\cal{O}}(10^{-(9-8)})$) for $m_{\tilde{B}} \lesssim 300$ GeV.
Therefore, an observation at the Brookhaven experiment of a deviation from
the SM prediction for $g_{\mu} - 2$
could signal a radiative muon mass.
An $s$-channel resonant Higgs production 
could test the radiative muon mass interpretation of a  
deviation in $g_{\mu} - 2$, as a radiative $m_{\mu}$
also implies  that the Higgs production rate is enhanced.
However, because of the constraint $m_{\tilde{B}} \gtrsim 300$ GeV
and the smallness of the muon mass,
the mixing $\sin2\theta_{\tilde{\mu}}$ is bounded from above
\cite{paper}.
As a result, the enhancement of 
$\bar{y}_{\mu}$ cannot be maximal (see eq.~(\ref{vertex2}))
and is typically only a few percent. 
Nevertheless, efforts to precisely determine $r_{\mu}$ 
are strongly motivated in such a situation.

\section{Prospects}

It has been shown that soft Yukawa couplings, {\em i.e.},
radiative generation of fermion masses, is a logical possibility
in certain (non-minimal) supersymmetric frameworks.
Aside from low-energy implications
for $CP$ violation and magnetic
moments, such scenarios imply an enhancement of the Higgs - fermion
couplings ranging from  a few percent up to an order of magnitude
in comparison with the MSSM.
Hence, the next generation of lepton colliders, and in particular,
the muon collider, offer an opportunity to determine experimentally
whether fermion masses are generated radiatively. 

The production rate on the  $s$-channel Higgs resonance
of a fermion whose mass is generated
radiatively is enhanced in comparison with the MSSM
by $r_{f}^{2}$, or if $m_{\mu}$ is also radiative, by $r_{\mu}^{2}r_{f}^{2}$. 
The partial width $\Gamma_{h^{0} \rightarrow ff}$
is enhanced by $r_{f}^{2}$.
The enhancement can partially cancel out in branching ratios, so
partial widths or the ratios
$\Gamma_{h^{0} \rightarrow ff}/\Gamma_{h^{0} \rightarrow WW^{*}}$
are more sensitive observables.
However, an enhancement of the total width
can also indicate radiative fermion masses. 
The projected errors of $\sim 5\%,\, 10-20\%, 10\%$ in the determination 
of $\Gamma^{2}_{h^{0} \rightarrow \mu\mu}$, $\Gamma^{2}_{h^{0} \rightarrow
cc,\,  bb}$ and the total width \cite{higgs}, respectively,
suggest that such tests are feasible.
However, detailed studies are needed.  
Decay modes of the heavy Higgs bosons, the Higgsinos, and the
sfermions offer additional probes of 
such scenarios in high energy machines \cite{paper}.
For example, if the $b$ mass is radiative then one expects
enhancement of $\tilde{b}_{1} \rightarrow H\tilde{b}_{2}$.
Similarly, in the case of, {\em e.g.}, the $c$-quark, one would also observe
mixing (which vanishes in the MSSM) between the scalar charms.

We would like to acknowledge discussions with V. Berger,
J. F. Gunion, H. E. Haber, and W. J. Marciano.


\begin{references}
\bibitem{susy} See, for example, contributions by Hinchliffe, Kelly,
               Paige, and Porod.
\bibitem{sob} Cheng C.-H., Feng J.L., and Polonsky N.,
               hep-ph/9706438; {\em ibid.}, hep-ph/9706476;
               Pierce D., Nojiri M., and Yamada Y., hep-ph/9707244;
               Pierce D. and Thomas S., SU-ITP 97-24.
\bibitem{vb}  Cheng {\em et al.} \cite{sob}; Barger V., in these proceedings.
\bibitem{schannel} Barger V., Berger M.S., Gunion J.F., and Han T., 
                   {\it Phys. Rev. Lett.} {\bf 75}, 1462 (1995);
                   {\em ibid.}, {\it Phys. Rept.} {\bf 286}, 1 (1997).
\bibitem{higgs} See Gunion's contribution and a summary talk by Han.
\bibitem{rad}  del Aguila F., Dugan M., Grinstein B., Hall L.J., Ross G.G.,
                and West P., {\it Nucl. Phys.} {\bf B250}, 225  (1985);
                Banks T., {\it Nucl. Phys.} {\bf B303}, 172  (1988);
                Ma E., {\it Phys. Rev. D} {\bf 39}, 1922 (1989);
                Krasnikov V.A., {\it Phys. Lett. B} {\bf 302}, 59 (1993);
                Arkani-Hamed N., Cheng C.-H., and Hall L.J.,
                {\it Phys. Rev. D} {\bf 54}, 2242 (1996).
\bibitem{paper} Borzumati F.M., Farrar G.R., Polonsky N., and Thomas S., 
                {\it Soft Yukawa couplings in supersymmetric theories}, 
                ZU-TH 23/97, RU-97-56, SU-ITP 97-45.
\end{references}
\end{document}